\documentclass[aps,onecolumn,pra,showpacs,tightenlines]{revtex4}
\usepackage{amsmath,mathrsfs,amsbsy,color,graphicx,bm,amsthm,amsfonts}
\usepackage{units,bbm,times,dcolumn,amssymb,epsfig,float}
\usepackage[colorlinks,citecolor=blue,linkcolor=blue,hyperindex,dvipdfm]{hyperref}

\begin{document}

\title{The precision of parameter estimation for dephasing model under squeezed reservoir}
\author{Shao-xiong Wu$^{1,2}$, and Chang-shui Yu$^{1}$\footnote{quaninformation@sina.com}}

\affiliation{ $^{1}$School of Physics and Optoelectronic Technology, Dalian University
of Technology, Dalian 116024, China\\
$^{2}$School of Science, North University of China, Taiyuan 030051,
China }

\date{\today }
\begin{abstract}
We study the precision of parameter estimation for dephasing model under squeezed environment. We analytically calculate the dephasing factor $\gamma(t)$ and obtain the analytic quantum Fisher
information (QFI) for the amplitude parameter $\alpha$ and the phase
parameter $\phi$. It is shown that the QFI for the amplitude
parameter $\alpha$ is invariant in the whole process, while the QFI for the
phase parameter $\phi$ strongly depends on the reservoir squeezing. It is shown that the QFI can be
enhanced for appropriate squeeze parameters $r$ and $\theta$. Finally, we also investigate the
effects of temperature on the QFI.
\end{abstract}
\maketitle

\section{Introduction}
Parameter estimation is one of the most important ingredients in various fields in both the classical and quantum worlds such as quantum metrology \cite{Giovannetti06,Giovannetti11,Demkowicz-Dobrzanski12}, gravitational wave detection \cite{Vallisneri08} and so on. Quantum Fisher information (QFI) is located at the central position in the parameter estimation, since accompanied with the Cram\'{e}r-Rao inequality, it is closely related to the sensitivity of the parameter \cite{Helstrom76,Braunstein94}. The essential work performed by Caves shows that quantum systems can provide more sensitivity than the classical ones and beat the shot noise limit in principle \cite{Caves81}. In recent year, improving the estimation precision became a significant issue in both experimentally and theoretically. In the realistic physical process, the quantum system unavoidably interacts with the surrounding environments \cite{Breuer02},  so  the effects of environments on the precision of parameter estimation attract intensive interests and lots of related work is reported. The precision spectroscopy using entangled states was proposed in the presence of the Markovian \cite{Huelga97} or non-Markovian noise \cite{Chin12}. The general framework was given for the estimation of the ultimate precision limit in noisy quantum-enhanced metrology \cite{Escher11}.  The dynamics of QFI was studied under the critical environment  \cite{SunZ10}. Quantum-enhanced metrology for multiple phase estimation \cite{Humphreys13} with noise \cite{YueJD14} was  also reported. In addition, in the framework of relativity theory, the quantum metrology has also been investigated recently \cite{Ahmadi14,TianZ15,WangJ14}.
In particular, lots of work have focused on increasing the precision of parameter estimation using different positive or negative methods \cite{TanQS13,Demkowicz-Dobrzanski14,Anisimov10,Dur14,Pezze09,Hyllus12,Joo11}.
However, in almost all the work,  the environment is always treated as vacuum or thermal reservoir. As we know,  the squeezed reservoir has been realized experimentally and widely applied in the relevant fields \cite{Scully97,Drummond04}.

In this paper, we investigate the effects of reservoir squeezing on the estimation precision of the amplitude parameter $\alpha$ and the phase parameter $\phi$. The model considered here is a two-level system with an initial amplitude parameter $\alpha$ and an embedded phase parameter $\phi$ undergoes a squeezed reservoir subjected to a dephasing process \cite{Breuer02} and finally the quantum Fisher information (QFI) of the estimated parameter is detected.  For this model, we analytically calculate the dephasing factor $\gamma(t)$  and the QFIs of the amplitude parameter $\alpha$ and the phase parameter $\phi$, respectively. We find that the QFI of the amplitude parameter $\alpha$ does not change with the presence of the dephasing process, which implies the information encoded in the amplitude parameter $\alpha$ is robust against the dephasing \cite{ZhongW13,YaoY14}. But  the QFI of the phase parameter $\phi$ obviously depends on the reservoir squeeze parameters. It is found that  $r$ (the module of the squeeze parameter) always play the negative roles  in the preservation of the QFI. However, $\theta$ (the reference phase for the squeezed filed) can retard the negative influence of the reservoir squeezing. If the parameter $\theta$ is chosen appropriately, the QFI of the phase parameter $\phi$ even can be enhanced. Finally, we follow the similar procedure to study the squeezed thermal reservoir and reveal the effects of temperature on the QFI.

\section{Quantum Fisher information}

To begin with, we would like to give a brief introduction about the quantum Fisher information (QFI). For the quantum state $\rho_{\phi}$,
the QFI of the estimated parameter $\phi$ is given by
\begin{eqnarray}
\mathcal{F}_{\phi}=\mathrm{Tr}\left(\rho_{\phi}L_{\phi}^2\right),
\end{eqnarray}
where the symmetric logarithmic derivative $L_{\phi}$ is defined by $%
2\partial_{\phi}\rho_{\phi}=L_{\phi}\rho_{\phi}+\rho_{\phi}L_{\phi}$. Considering the eigenvalues $\lambda _{i}$ of the quantum state $\rho_{\phi}$ and the corresponding eigenvectors $%
\left\vert \varphi _{i}\right\rangle$,  the QFI can be explicitly given as \cite{ZhangYM13,MaJ11}
\begin{eqnarray}
\mathcal{F}_{\phi }=\sum_{i}\frac{(\partial _{\phi }\lambda _{i})^{2}}{\lambda
_{i}}+\sum_{i\neq j}\frac{2(\lambda _{i}-\lambda _{j})^{2}}{%
\lambda_{i}+\lambda _{j}}\left\vert \left\langle \varphi _{i}|\partial
_{\phi}\varphi _{j}\right\rangle \right\vert ^{2}  \label{fisher1}
\end{eqnarray}
with $\vert\partial_{\phi}\varphi\rangle$ denoting the partial derivative $%
\partial\vert\varphi\rangle/\partial{\phi}$. For pure state $\vert\varphi\rangle$, the QFI can be
given by a more simple expression $\mathcal{F}_{\phi
}=4(\left\langle \partial _{\phi }\varphi |\partial _{\phi
}\varphi\right\rangle -\left\vert \left\langle \partial _{\phi }\varphi
|\varphi\right\rangle \right\vert ^{2})$.

Based on the QFI and  the quantum Cram\'{e}r-Rao inequality, one can find that the precision $\delta$ of the parameter $\phi$ can be expressed   as \cite{Helstrom76,Braunstein94}
\begin{eqnarray}
\delta(\phi)\geq\frac{1}{\sqrt{\nu \mathcal{F}_{\phi}}},  \label{CRinequality}
\end{eqnarray}
where $%
\nu$ is the number of repeated experiments. So the larger value of the QFI implies the higher
sensitivity of the estimated parameter $\phi$. This shows the importance of
the QFI in parameter estimation.

\section{The Hamiltonian and the evolution under the squeezed vacuum reservoir}

We assume that the input state is a two-level superposition state $\left\vert \varphi _{\alpha}\right\rangle=\cos\frac{\alpha}{2}\left\vert e\right\rangle+\sin\frac{\alpha}{2}\left\vert g\right\rangle$. Before going through the dephasing process, a phase gate $U_{\phi }=\left\vert g\right\rangle\left\langle g\right\vert +e^{i\phi }\left\vert e\right\rangle \left\langle e\right\vert$ is operated on the input state $|\varphi_{\alpha}\rangle $. So, the output state is given by
\begin{eqnarray}
\rho _{\mathrm{out}}=U_{\phi }\left\vert \varphi _{\alpha}\right\rangle
\left\langle \varphi _{\alpha}\right\vert U_{\phi }^{\dagger }.
\label{rhoout}
\end{eqnarray}
The output state $\rho _{\mathrm{out}}$ contains two parameters: $\alpha$ (we call it the amplitude parameter) and $\phi$ (the phase parameter). Then the state $\rho _{\mathrm{out}}$  couples to the environment and undergoes a dephasing process. The environment $\rho _{\mathrm{bath}}$ is assumed to be the squeezed vacuum reservoir, which is given by
\begin{eqnarray}
\rho _{\mathrm{bath}}=\prod_{k}S_{k}\left( r,\theta \right) \left\vert
0_{k}\right\rangle \left\langle 0_{k}\right\vert S_{k}\left( r,\theta
\right) ^{\dagger }.\label{squeezedvacuum}
\end{eqnarray}
The unitary squeeze operator $S_{k}\left( r,\theta \right) $ is given by
\begin{eqnarray}
S_{k}\left( r,\theta \right) =\exp \left( \frac{1}{2}re^{-i\theta }b_{k}^{2}-%
\frac{1}{2}re^{i\theta }b_{k}^{\dagger 2}\right),\label{squeezedoperator}
\end{eqnarray}
where $r$ is the squeeze parameter and $\theta $ is the reference phase.

The total Hamiltonian for the system plus environment is given by
\begin{eqnarray}
H&=&H_{0}+H_{\mathrm{I}}
\end{eqnarray}
with
\begin{eqnarray}
H_{0} &=&\frac{\omega _{0}}{2}\sigma _{z}+\sum_{k}\omega _{k}b_{k}^{\dagger
}b_{k},  \notag \\
H_{\mathrm{I}} &=&\sum_{k}g_{k}\sigma _{z}\big(b_{k}+b_{k}^{\dagger }\big) ,
\end{eqnarray}%
where $\omega _{0}$ is the transition frequency between the two levels, $\omega _{k}$ is the frequency of the $k$-th reservoir mode, $b_{k}(b_{k}^{\dagger })$ is the annihilation (creation) operator and $g_{k}$ is the coupling constant between the system and the environment. The initial state of the system plus environment is a product state $\rho_{\rm{out}}\otimes\rho_{\rm{bath}}$. After a standard calculation given in Appendix, one can obtain that the final state after the dephasing processing is
\begin{eqnarray}
\rho _{\mathrm{s}}(t)=\left[
\begin{array}{cc}
\cos ^{2}\frac{\alpha }{2} & \frac{1}{2}e^{-\gamma (t)+i\phi }\sin \alpha \\
\frac{1}{2}e^{-\gamma (t)-i\phi }\sin \alpha & \sin ^{2}\frac{\alpha }{2}%
\end{array}%
\right] , \label{rhos}
\end{eqnarray}%
where the dephasing factor $\gamma (t)$ is given by
\begin{eqnarray}
\gamma (t)=\int d\omega J(\omega )\frac{1-\cos (\omega t)}{\omega ^{2}}\big[%
\cosh (2r)-\cos (\omega t-\theta )\sinh (2r)\big] \label{dephasingfactor}
\end{eqnarray}%
with  $J(\omega) $ denoting the spectral density of the environment.

In order to give a concrete example of the parameter estimation scheme, we consider the structure of the environment is the Ohmic-like spectrum with soft cutoff \cite{Breuer02,Addis14}
\begin{eqnarray}
J(\omega )=\eta \frac{\omega ^{s}}{\omega _{c}^{s-1}}\exp \left(-\frac{%
\omega }{\omega _{c}}\right),
\end{eqnarray}
where $\omega _{c}$ is the high frequency cutoff, $\eta $ is the dimensionless coupling constant. The parameter $s$ is positive and determines the property of the environment. For $s<1$, the environment is the sub-Ohmic reservoir; for $s=1$, the environment is the Ohmic reservoir; and for $s>1$, the environment is the super-Ohmic reservoir. For the sake of simplicity, we will assume the cutoff frequency $\omega _{c}$ is $1$ in the rest of this paper. After some algebra, one can obtain that the dephasing factor $\gamma(t)$ is given by
\begin{eqnarray}
\gamma (t)&=&\frac{\eta }{4}\Gamma (s-1)\Big\{2\cosh (2r)[2-(1+it)^{1-s}-(1-it)^{1-s}]
\notag \\
&&+e^{-i\theta }\sinh (2r)\left[ 1-2(1-it)^{1-s}+(1-2it)^{1-s}\right]  \notag
\\
&&+e^{i\theta }\sinh (2r)\left[ 1-2(1+it)^{1-s}+(1+2it)^{1-s}\right] \Big\},\notag\\
\label{gammat1}
\end{eqnarray}
where $\Gamma(\cdot)$ is the Euler Gamma function.

Substituting the estimated state $\rho _{\mathrm{s}}(t)$ (Eq. (\ref{rhos})) into the formula of QFI (Eq. (\ref{fisher1})), the analytic expression for the QFIs of the amplitude parameter $\alpha $ and the phase parameter $\phi $ can be given as \cite{ZhongW13,YaoY14}
\begin{eqnarray}
\mathcal{F}_{\alpha } &=&1, \\
\mathcal{F}_{\phi } &=&e^{-2\gamma (t)}.
\end{eqnarray}
It is easy to find that the QFI of the amplitude parameter $\alpha$ does not affected by the dephasing factor $\gamma(t)$ and keeps the constant $1$ in the dephasing process. It implies that the information encoded in the $\alpha$ is immune to the environment in this parameter estimation scheme \cite{ZhongW13,YaoY14}. The QFI of the phase parameter $\phi$ can be influenced by the dephasing factor $\gamma(t)$ and independent of the value of the estimated parameter $\phi$. In the following, we will investigate the effects of the reservoir squeezing on the QFI of the phase parameter $\phi$.

\section{The effects of reservoir squeezing on the QFI}
\begin{figure}[h!]
  \centering
  \includegraphics[width=0.5\columnwidth]{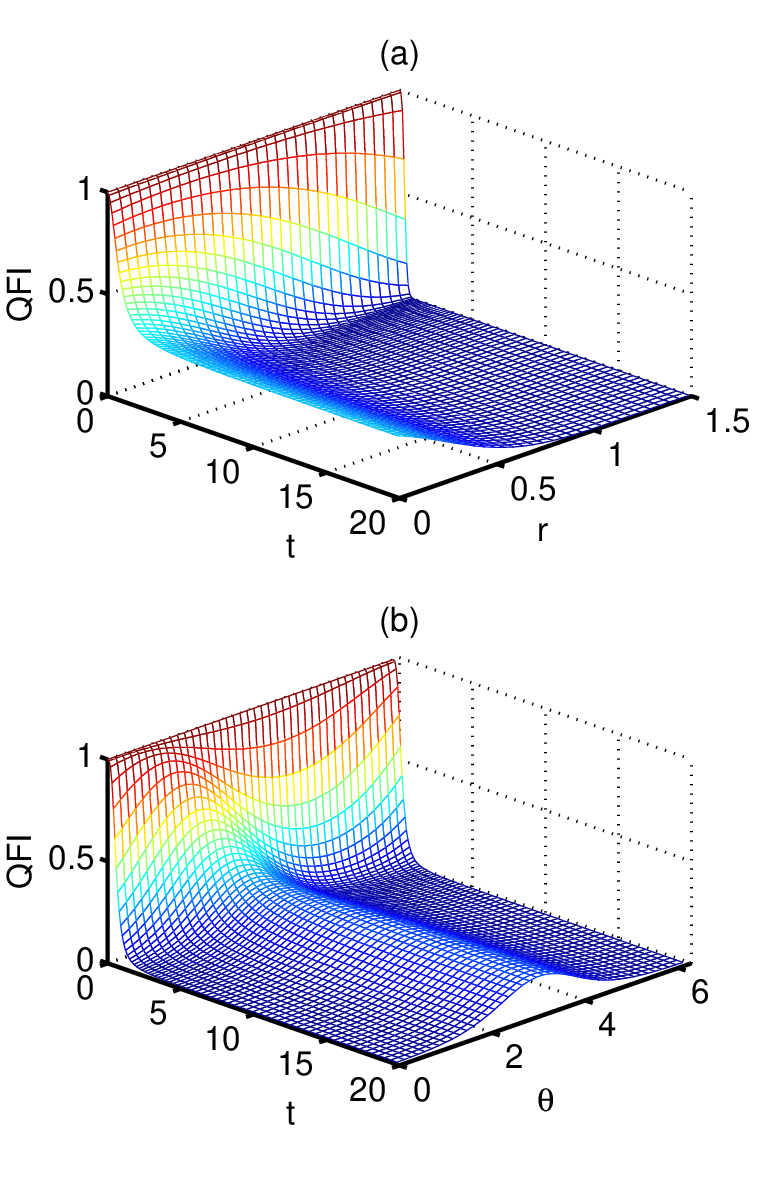}\\
  \caption{ QFI under squeezed vacuum reservoir. QFI of the phase parameter $\phi$ vs. the parameters $r $, $\theta$ and $s$. The reservoir is the super-Ohmic environment $s=2$, and the coupling constant between the system and the environment is $\eta=0.6$. In Panel (a), the phase parameter for the squeezed field is chosen $\theta=0$. The squeeze parameter is $r=0.8$ in Panel (b).}\label{yasuo1}
\end{figure}
In order to compare the effects of the reservoir squeezing on the precision of the parameter estimation, we will first study the case that the reservoir is not squeezed, i.e., the vacuum reservoir. The dephasing factor $\gamma_0(t)$ in the vacuum reservoir can be simplified as follows
\begin{eqnarray}
\gamma_0 (t) &=&\int d\omega J\left( \omega \right) \frac{1-\cos \omega t}{%
\omega ^{2}}  \notag \\
&=&\eta \left[ 1-\frac{\cos [(s-1)\arctan (t)]}{(1+t^{2})^{(s-1)/2}}\right]
\Gamma (s-1).  \label{gammatvacuum}
\end{eqnarray}

For the squeezed vacuum reservoir, the dephasing factor $\gamma(t)$ (Eq. (\ref{gammat1})) involves not only the Ohmic parameter $s$ but also the squeeze parameters $r$ and $\theta$. Compared with the vacuum reservoir, one can easily find that both the squeeze parameters $r$ and $\theta$ take effects,  the competition between the squeeze parameters $r$ and $\theta$ determines the behavior of the QFI's dynamics. The dynamics of the QFI with the parameters $r$, $\theta$ and $s$ are drawn in Fig. \ref{yasuo1}. In Panel (a), the parameter $s$ is chosen as the super-Ohmic environment $s=2$ and the reference phase for the squeezed field is fixed at $\theta=0$. One can find that the squeeze parameter $r$ always plays the negative roles in the dynamics of QFI, that is, the larger reservoir squeezing the lower value of QFI. Even though the reservoir squeezing $r$ is harmful to preservation of the QFI, the phase parameter $\theta$ can retard the negative effect induced by the reservoir squeezing, and even eliminate its influence, which can be concluded from Panel (b). In Panel (b), the reservoir is also chosen as the super-Ohmic environment $s=2$ and the squeeze parameter is chosen as $r=0.8$. One can find that, the QFI can be preserved larger and longer when $\theta$ is in vicinity of $\pi$. In this sense, if the squeeze parameters $r$ and $\theta$ are chosen properly, the QFI even can be enhanced in the dephasing process.

\begin{figure}[h!]
  \centering
  \includegraphics[width=0.5\columnwidth]{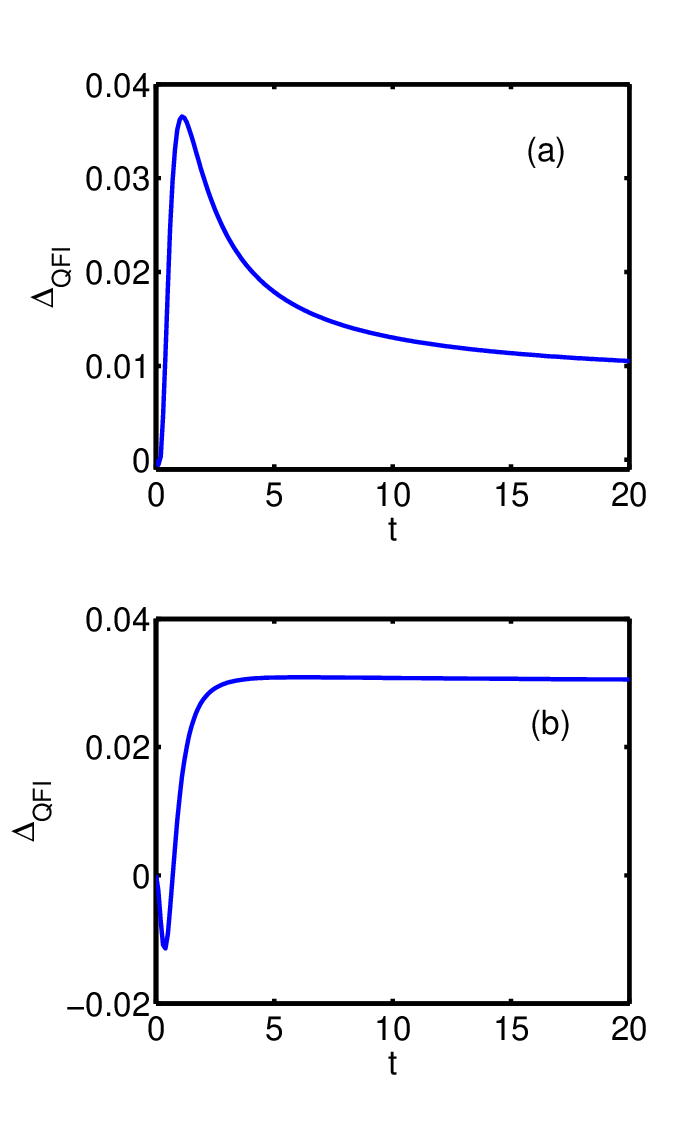}\\
  \caption{The differences of QFIs for vacuum reservoir with and without reservoir squeezing. In Panel (a), the phase parameter for the squeezed field is $\theta=2$. In Panel (b), the phase parameter for the squeezed field is $\theta=3$. In both Panels, the reservoir is under super-Ohmic environment $s=2$, the reservoir squeeze parameter is $r=0.1$ and the coupling constant is $\eta=0.6$.}\label{yasuo2}
\end{figure}

In Fig. \ref{yasuo2}, we plot the difference between the QFI under the squeezed vacuum reservoir and the vacuum reservoir. For both panels, the reservoir is super-Ohmic environment $s=2$, the squeeze parameter is chosen as $r=0.1$. In Panel (a), the phase parameter for the squeezed field is chosen as $\theta=2$. Compared with the vacuum reservoir, the dynamics of QFI under the squeezed vacuum reservoir can always be enhanced, especially when $t$ is in the vicinity of $1$. In Panel (b), the phase parameter for the squeezed field is chosen as $\theta=3$. Comparing the squeezed vacuum reservoir with the vacuum reservoir, we can find that the QFI is descended at first, but the QFI can be enhanced in the later time. This shows that the squeeze phase parameter $\theta$ plays more significant role than the squeeze parameter $r$ in this parameter estimation scheme. The competition between the squeeze parameter $r$ and the squeeze phase parameter $\theta$ determines whether the QFI is enhanced or descended in the dephasing process. In Figs. \ref{yasuo1} and \ref{yasuo2}, the coupling constant parameter is chosen as $\eta=0.6$. In the Ohmic reservoir and the sub-Ohmic reservoir, one can find the similar property that the QFI can be improved when the parameters $r$ and $\theta$ are chosen appropriately.

\section{The effects of the temperature on the QFI}

We emphasize that the above parameter estimation scheme can also be used to investigate the effects of the temperature on the QFI, i.e., the considered environment is a squeezed thermal reservoir
\begin{eqnarray}
\rho _{\mathrm{bath}}=\prod_{k}S_{k}(r,\theta )\rho _{\mathrm{th}%
}S_{k}(r,\theta )^{\dagger }.
\end{eqnarray}
The squeeze operator $S_{k}(r,\theta )$ is given in Eq. (\ref{squeezedoperator}) and the thermal state is $\rho _{\mathrm{th}}=\frac{\exp(-\beta \omega _{k}b_{k}^{\dagger }b_{k})}{\mathrm{Tr}\exp (-\beta \omega_{k}b_{k}^{\dagger }b_{k})}$. Here, the parameter $\beta =1/(kT)$, $k$ and $T$ denotes the Boltzman constant and the temperature, respectively. If the temperature is $0$, the thermal state will become the vacuum state $|0_k\rangle\langle 0_k|$, and the environment will become the squeezed vacuum reservoir, which is given in Eq. (\ref{squeezedvacuum}).

Following the derivation in Appendix, we can obtain the dephasing factor $\gamma_{T}(t)$ for the squeezed thermal reservoir can be given by
\begin{eqnarray}
\gamma_{\mathrm{T}}(t)=(2\langle n\rangle+1)\gamma(t),  \label{gammatt}
\end{eqnarray}
with $\langle n\rangle=1/(\exp(\beta\omega_0)-1)$, and $\gamma(t)$ is the dephasing factor for the squeezed vacuum reservoir, which is given in Eq. (\ref{gammat1}). Therefore, the dynamics of the diagonal elements of the reduced density matrix remain invariant, the off-diagonal element of the reduced density matrix is determined by
\begin{eqnarray}
\rho_{10}(t)=\exp[-\gamma_{\mathrm{T}}(t)] =\rho_{01}(t)^*.
\end{eqnarray}

\begin{figure}[h!]
  \centering
  \includegraphics[width=0.5\columnwidth]{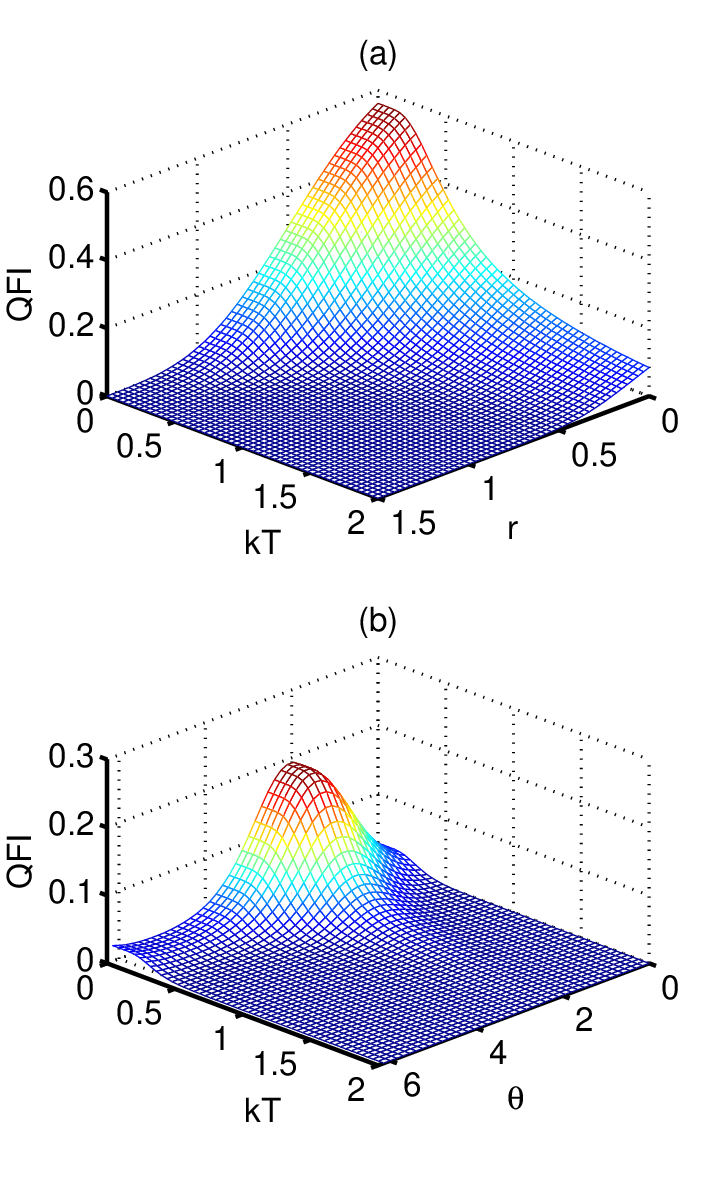}
  \caption{QFI under the squeezed thermal reservoir. The evolution time $t$ is chosen as $t=5$. The coupling constant $\eta=0.6$. In Panel (a), the environment is super-Ohmic with $s=2$ and $\theta=0$. In Panel (b), the environment is sub-Ohmic with $s=0.8$ with $r=0.8$.}\label{yasuo_thermal}
  \end{figure}

In Fig. \ref{yasuo_thermal}, we plot the dynamics of the QFI under different temperature $T$. One can easily find that the high temperature  always enhances the decay of the QFI. This can be explained easily. The high temperature always accelerate the dephasing, which can be seen from Eq. (\ref{gammatt}).
\section{Conclusion and discussions}

In this paper, we investigated the effects of the reservoir squeezing on the precision of parameter estimation (the amplitude parameter $\alpha$ and the phase parameter $\phi$) in the dephasing process subjected to the squeezed vacuum (and thermal) reservoir. For the amplitude parameter $\alpha$, the QFI does not change in the dephasing process. However, the QFI of the phase parameter $\phi$ is affected by the reservoir squeezing. The squeeze parameter $r$ always plays the negative role in the dynamics for QFI of $\phi$, while the phase parameter for the squeezed field $\theta$ can retard/weaken this influence, even enhance the QFI when the parameter $\theta$ is chosen appropriately. In the end, we also investigate the effects of temperature on the QFI.

The physical model in this paper maybe realized by Bose-Einstein Condensates using dephasing collisions \cite{Bar-Gill11}. At last, we would like to say that the parameter estimation of multiple phase  \cite{Humphreys13,YueJD14} or initially $N$ entangled Greenberg-Horne-Zeilinger state for the dephasing model under squeezed reservoir are deserved our further investigation.

\section* {Acknowledgement}

This work was supported by the National Natural Science Foundation of China, under Grant No.11375036 and 11175033, the Xinghai Scholar Cultivation Plan and the Fundamental Research Funds for the Central Universities under Grant No. DUT15LK35.

\renewcommand{\theequation}{A\arabic{equation}}
\setcounter{equation}{0}
\section*{Appendix:  derive the dephasing factor $\gamma (t)$ }

In this section, we will derive the dephasing factor $\gamma (t)$ (Eq. (\ref{gammat1})) in the dephasing processing under the squeezed vacuum reservoir. In the interaction picture, the interaction Hamiltonian can be written as
\begin{eqnarray}
H_{\mathrm{I}}(t) &=&e^{iH_{0}t}H_{\mathrm{I}}e^{-iH_{0}t}  \notag \\
&=&\sum_{k}g_{k}\sigma _{z}\big(b_{k}^{\dagger }e^{i\omega
_{k}t}+b_{k}e^{-i\omega _{k}t}\big).
\end{eqnarray}%
According to the Magnus expression \cite{Blanes10}, the unitary time-evolution operator in the interaction picture can be given by
\begin{eqnarray}
U(t) &=&\mathrm{T}_{\leftarrow }\exp \left[ -i\int_{0}^{t}dt^{\prime }H_{%
\mathrm{I}}(t^{\prime })\right]  \notag \\
&=&C(t)\cdot V(t),  \label{unitarytimeevolution}
\end{eqnarray}%
where the time-dependent complex number $C(t)$ is given by $C(t)=\exp \left[\sum_{k}g_{k}^{2}\frac{\sin (\omega _{k}t)-\omega _{k}t}{i\omega _{k}^{2}}\right] $, and the unitary operator $V(t)$ is defined as
$V(t)=\exp \left[ \sigma _{z}\sum_{k}\left( \alpha _{k}b_{k}^{\dagger
}-\alpha _{k}^{\ast }b_{k}\right) \right]$ with the amplitude coefficient $\alpha _{k}=g_{k}\frac{1-e^{i\omega _{k}t}}{\omega _{k}}$.

The initial state of the system plus the environment is assumed to be a product state $\rho _{\mathrm{s}}(0)\otimes \rho _{\mathrm{bath}}$. Due to the communication between the system's Hamiltonian $\frac{\sigma_z\omega_k}{2}$ and the interaction Hamiltonian $\sum_kg_k\sigma_z(b_k+b_k^{\dagger})$, the evolution of the reduced density matrix element for dephasing processing under squeezed vacuum reservoir is governed by \cite{Breuer02}
\begin{eqnarray}
\rho _{ij}\left( t\right)=\left\langle i\right\vert \mathrm{Tr}_{\mathrm{%
bath}}\left\{ V(t)\rho _{s}(0)\otimes \rho _{\mathrm{bath}%
}V^{\dagger}(t)\right\} \left\vert j\right\rangle.  \label{rhoij}
\end{eqnarray}%
In the above equation, the time-dependent complex number $C(t)$ multiplied by its complex conjugation is equal to unit, so it can be omitted. For the diagonal elements of the reduced density matrix, it is easy to prove that the elements do not evolute, i.e., $\rho _{11}(t)=\rho_{11}(0)$ and $\rho_{00}(t)=\rho _{00}(0)$.

Then, we will characterize the dynamics of the off-diagonal elements going through the dephasing process under the squeezed vacuum reservoir. Substituting $V(t)$ into Eq. (\ref{rhoij}), the evolution of the element $\rho _{10}\left( t\right) $ can be given as follows
\begin{eqnarray}
\rho _{10}\left( t\right)  &=&\mathrm{Tr}_{\mathrm{bath}}\{ \exp [
\sum_{k}2( \alpha _{k}b_{k}^{\dagger }-\alpha _{k}^{\ast }b_{k}) ] \rho _{\mathrm{bath}}\} \rho _{10}( 0)   \notag \\
&=&\prod_{k^{\prime }}\langle 0_{k^{\prime }}\vert S_{k^{\prime
}}^{\dagger }\exp [ \sum_{k}2( \alpha _{k}b_{k}^{\dagger }-\alpha
_{k}^{\ast }b_{k}) ] S_{k^{\prime }}\vert 0_{k^{\prime
}}\rangle \rho _{10}( 0)   \notag \\
&=&\exp [ -\sum_{k}2\vert \beta _{k}\vert ^{2}] \rho
_{10}( 0),  \label{rho10t}
\end{eqnarray}%
where the complex number coefficient $\beta _{k}=\alpha _{k}\cosh r+e^{i\theta }\alpha _{k}^{\ast }\sinh r$.

We can define the parameter $\sum_{k}2\left\vert \beta _{k}\right\vert ^{2}$ being the dephasing factor $\gamma (t)$,
\begin{eqnarray}
\gamma (t) \sum_{k}\left\vert \alpha _{k}\right\vert ^{2}\left[ 2\cosh (2r)+\left(
\alpha _{k}^{2}e^{-i\theta }+\alpha _{k}^{\ast 2}e^{i\theta }\right) \sinh
(2r)\right] . \label{gammat}
\end{eqnarray}
So the evolution of the element $\rho _{10}\left( t\right) $ is governed by
\begin{eqnarray}
\rho _{10}(t) =\exp [-\gamma (t)]\rho _{10}(0) =\rho _{01}(t)^{\ast }.
\end{eqnarray}%
Substituting the amplitude coefficient $\alpha _{k}$ in the time-evolution operator $U(t)$ into the dephasing factor $\gamma(t) $, we can obtain the analytic expression for the dephasing factor as
\begin{eqnarray}
\gamma (t)=\sum_{k}4g_{k}^{2}\frac{1-\cos \omega _{k}t}{\omega _{k}^{2}}%
\left[ \cosh (2r)-\cos (\omega _{k}t-\theta )\sinh (2r)\right].
\label{dephasingfactor_appendix}
\end{eqnarray}
Considering the spectrum density $J(\omega )$ of the modes for the frequency $%
\omega $ \cite{Breuer02} as $J(\omega )=4f(\omega )|g(\omega)|^{2}$, performing the continuum limit of the reservoir modes and changing the sum on $k$ in Eq. (\ref{dephasingfactor_appendix}) to the integral on  the frequency $\omega$, we can obtain the dephasing factor $\gamma(t)$ in Eq. (\ref{dephasingfactor}).

\end{document}